\documentclass[letterpaper, 10 pt, conference]{ieeeconf}  
\IEEEoverridecommandlockouts                              
\usepackage[utf8]{inputenc}
\usepackage[T1]{fontenc}
\usepackage{graphicx}
\usepackage{hyperref}
\usepackage{url}
\usepackage[english]{babel}
\usepackage{adjustbox}
\usepackage{tabularx}
\usepackage{subcaption}
\usepackage[rawfloats=true]{floatrow} 
\restylefloat{figure}

\usepackage{tabularx} 

\title{\LARGE \bf
Evolving Performance Practices in Beethoven’s Cello
Sonatas: Tempo, Portamento, and Historical
Interpretation of the First Movements}

\author{Ignasi Sole \hspace{0.5cm} \texttt{isolepinas@gmail.com} \hspace{0.5cm} \today}

\makeatletter
\renewcommand{\@biblabel}[1]{[#1]} 
\renewcommand{\thebibliography}[1]{%
  \section*{\textbf{References}} 
  \small 
  \list{\@biblabel{\@arabic\c@enumiv}}{%
    \settowidth\labelwidth{\@biblabel{#1}}%
    \leftmargin\labelwidth
    \advance\leftmargin\labelsep
    \usecounter{enumiv}%
    \let\p@enumiv\@empty
    \renewcommand\theenumiv{\@arabic\c@enumiv}}%
  \sloppy\clubpenalty4000\widowpenalty4000%
  \sfcode`\.\@m}
\makeatother
\begin{document}

\maketitle
\thispagestyle{empty}
\pagestyle{empty}



\begin{abstract}

This paper examines the evolving performance practices of Ludwig van Beethoven's cello sonatas, with a particular focus on tempo and portamento between 1930 and 2012. It integrates analyses of 22 historical recordings, advancements in recording technology to shed light on changes in interpretative approaches. By comparing Beethoven's metronome markings, as understood through contemporaries such as Czerny and Moscheles, with their application in modern performances, my research highlights notable deviations. These differences prove the challenges performers face in reconciling historical tempos with the demands of contemporary performance practice. My study pays special attention to the diminishing use of audible portamento in the latter half of the 20th century, contrasted with a gradual increase in tempo after 1970. This development is linked to broader cultural and pedagogical shifts, including the adoption of fingering techniques that reduce hand shifts, thereby facilitating greater technical precision at faster tempos. Nonetheless, my study identifies the persistence of “silent portamento” as an expressive device, allowing performers to retain stylistic expression without compromising rhythmic integrity. My paper offers valuable insights for performers and scholars alike, advocating a critical reassessment of Beethoven's tempo markings and the nuanced application of portamento in modern performance practice.

\end{abstract}

\section{Methodologies}

To achieve the results discussed in this paper, I employed a combination of audio analysis tools and techniques to investigate tempo, time, and portamento in Beethoven’s sonatas for piano and cello. These methodologies integrate traditional musicological approaches with computational tools, providing a detailed and multifaceted analysis.

Tempo and time were analysed using a dual approach. Performance recordings from 1930 to 2012 were segmented into sections and subsections. Beats-per-minute (BPM) were calculated for individual bars and averaged across sections, movements, and entire sonatas. This data was visualised through scatter plots, histograms, and tempographs, revealing trends in tempo evolution over time. Tools such as Audacity and Sonic Visualizer facilitated the extraction of BPM and duration metrics. Sonic Visualizer’s spectrogram functionality enabled detailed temporal and spectral analysis, while its measurement tools allowed high-resolution timing analysis for specific passages.

For portamento analysis, an innovative methodology was developed to address the challenges of polyphonic recordings. While previous studies relied on qualitative annotations and small-scale analyses, this study introduced a quantitative approach incorporating two variables: sliding and silent portamento. Both were identified manually through auditory analysis and visualised using Sonic Visualizer’s melodic spectrogram layer.

To measure portamento length and steepness, spectrogram data was imported into the GNU Image Manipulation Program (GIMP). The slope of portamento slides was calculated using a gradient formula, enabling quantification of the speed and character of portamento transitions. Adjustments to spectrogram visibility and gain were necessary for older recordings, where spectral clarity was diminished due to analog-to-digital conversion artifacts.

Annotated scores were transferred to Sibelius, where portamento occurrences were marked with symbols for sliding (wavy lines) and silent (dotted lines) portamento. This transcription enabled systematic comparison of portamento frequency and location across recordings. Data analysis using scatter plots with linear regressions assessed trends over time, revealing a decline in sliding portamento coinciding with faster tempos and shifts in cello technique. Silent portamento emerged as a subtle alternative to maintain expressive transitions. In the following section I will introduce two hypothesis which discuss the reasons behind the fastness of the metronome marks that Beethoven left behind and how unattainable they are for the performance of Beethoven's cello sonatas.

\section{Beethoven's Metronome}

The tempos in Beethoven’s cello sonatas have sparked significant scholarly debate due to the contrast between Beethoven's metronome markings and practical performance speeds. Multiple theories attempt to explain this disparity, ranging from mechanical issues with Beethoven's metronome to the influence of piano design.

Beethoven's metronome (one of the earliest designs by Johann Nepomuk Maelzel (1772–1838) who was a German inventor best known for designing one of the first practical metronomes, which standardised tempo indications for musicians), faced documented issues. Letters reveal Beethoven’s complaints about the metronome’s ``uneven pulse'' and frequent repairs, raising questions about its accuracy.~\cite{c1} Musicologists Lawrence Talbot and Sture Forsen hypothesised that mechanical degradation or misalignment in the pendulum could have caused Beethoven to set tempos that were unintentionally fast.~\cite{c2}\cite{c3} Additionally, researchers Iñaki Ucar and Almudena Martín-Castro propose that Beethoven may have misread the triangular measurement indicator of the metronome, potentially adding an unintentional 12 BPM to his markings.~\cite{c4}

\begin{figure}
    \centering
    \includegraphics[width=0.5\linewidth]{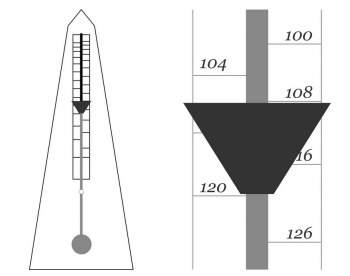}
    \caption{Example of original metronome design, which Beethoven could have misread.}
    \label{fig:enter-label}
\end{figure}


The design of Beethoven's pianos significantly influenced tempo feasibility. Fortepianos in Beethoven’s era had rapid tone decay due to leather damping mechanisms, which necessitated faster tempos to sustain musical continuity. Steven Isserlis notes that the fortepiano’s quick decay likely compelled performers to adopt faster tempos than those used on modern pianos, which have a longer sustain due to steel strings and felt-covered hammers.~\cite{c5} Conversely, the heavier action of pianos like the Erard posed challenges for quick passages. Beethoven briefly favored the Erard piano for its extended tonal range but later returned to lighter-action Viennese pianos like those of Stein, which facilitated faster tempos.~\cite{c6} This adaptability in instrument design influenced the development of Beethoven’s tempo indications, as lighter pianos allowed for greater speed, while heavier instruments necessitated slower tempos.

Modern performances frequently fall short of Beethoven's marked tempos, with differences of up to 40 BPM in some movements.~\cite{c7} While some scholars argue these discrepancies reflect inherent limitations of modern performers, others point to changing cultural and aesthetic values. For example, lyrical sonatas like Op. 69 have maintained slower tempos across decades, as performers prioritize expressiveness over adhering to strict tempo markings. Notably, advancements in piano design have indirectly influenced tempo choices. The increased sustain and resonance of modern pianos allow performers to adopt slower tempos without risking the loss of musical continuity. This contrasts with the fortepiano's tonal decay, which naturally encouraged faster tempos to avoid empty spaces in the sound.

\section{Tempo in Beethoven's Cello Sonatas}

Noorduin's analysis of Beethoven's metronome markings highlights the significant challenges performers face in achieving the specified tempos.~\cite{c7} By examining the markings attributed to Beethoven, Carl Czerny, and Ignaz Moscheles, alongside those proposed by Rudolf Kolisch in 1993, my study investigates the tempo trends in recorded performances of Beethoven's cello sonatas. Kolisch, a Viennese violinist and advocate of historically informed performance practices, proposed alternative metronome markings for Beethoven’s works, including his cello sonatas. His suggestions reflect an effort to interpret Beethoven’s intended tempos in a manner that balances historical context with practical considerations for modern performance.

MY analysis focuses on 22 recordings, divided into two chronological groups: 1930–1970 and 1970–2012. The division reflects significant shifts in performance practices, with the latter period encompassing the emergence and widespread influence of the Historically Informed Performance (HIP) movement. This division allows for a clearer analysis of trends in tempo evolution and their relationship to broader changes in interpretive approaches.

\begin{table}[htp!]
\centering
\caption[Beethoven's metronome marks for the cello sonatas according to different sources]{Beethoven's metronome marks for the cello sonatas according to different sources.}

\begin{tabular}{ c c c c c c p{5cm}}
\hline
\hline
Sonata & Ko. & Mo. & Cz. & 1930-1970 & 1970-2012 \\ 
 \hline
Op. 5 No. 1   & 126         & 160          & 160    &   135.78  & 140.98  \\
Op. 5 No. 2   & 126         & 252           & 252    &   214.43   & 215.22 \\
Op. 69        & 144         & 144          & 144     & 127.63    &   127.45  \\
Op. 102 No. 1 & 144         & 176          & 152    & 135.52      &  140.71  \\
Op. 102 No. 2 & n/a       & 168            & 152      &   123.83      &  126.05 \\
\hline
\end{tabular}
\end{table}

Table 1 summarises the metronome markings from historical sources compared to the average tempos in recorded performances. Across all sonatas, the recorded tempos consistently fall below the prescribed values, underscoring the technical challenges of realising Beethoven’s tempo indications. For example, Op. 5 No. 2 demonstrates a marked discrepancy, with Moscheles and Czerny suggesting a tempo of 252, while the recorded averages are 214.43 (1930–1970) and 215.22 (1970–2012).

A discernible trend of tempo acceleration is observed in performances post-1970, particularly in sonatas such as Op. 5 No. 1 (135.78 to 140.98) and Op. 102 No. 1 (135.52 to 140.71). This increase corresponds with the HIP movement’s advocacy for faster and more agile interpretations. As I mentioned earlier, Op. 69 is a special case, showing minimal variation in tempo (127.63 to 127.45). Its lyrical and expressive qualities appear to have guided a consistent interpretive approach across generations.

My findings indicate that the tempo markings provided by Czerny, Moscheles, and Kolisch may represent theoretical upper limits rather than practical guidelines for performance. This aligns with the broader observation that modern performers strive to balance historical ideals with contemporary technical and interpretive constraints. However, this raises the question: how do tempo changes influence the broader style and interpretation of these sonatas?

Among various aspects of performance style, portamento stands out as a particularly tangible and expressive feature that is closely tied to tempo. Changes in tempo can significantly shape the use and effectiveness of portamento, making it a logical focus for deeper exploration. By examining portamento, I aim to uncover how performers negotiate tempo-related challenges while maintaining the expressive and technical demands of Beethoven’s music. Ultimately changes in expression that are forcefully driven by changes in speed, are effectively changes in performance practise and tradition.

\section{Portamento in Beethoven's Cello Sonatas}

Portamento can be categorised into two forms: sliding portamento, where the finger slides audibly between notes to create a melodic connection, and silent portamento, where the slide itself is inaudible but a faint metallic sound from the finger gliding across the string can still be detected. I propose silent portamento as new variable to take into consideration when analysing expression in performance practise.

The analysis reveals a consistent decline in sliding portamento usage across Beethoven’s cello sonatas from 1930 to 2012. Measured reductions range from 34.7\% in Op. 69 to 61.09\% in Op. 5 No. 2. These decreases occur irrespective of tempo or performance duration, suggesting that the decline reflects broader interpretive and stylistic trends rather than technical constraints. Silent portamento, however, shows inconsistent trends: increases of 43.3\% and 163.6\% are observed in Op. 69 and Op. 102 No. 1, while Op. 5 No. 1 and Op. 5 No. 2 show reductions of 29.1\% and 11.11\%, respectively. This variability suggests that silent portamento is not a direct replacement for sliding portamento and is used contextually.

The relationship between tempo and portamento is complex. While faster tempos may limit the feasibility of sliding portamento, tempo alone does not dictate its usage. For example, in Op. 5 No. 1, Casals (1954) performed at 105.83 bpm with 20 sliding portamentos, while Maisky (1990–92) performed at 122.51 bpm with only 6 sliding portamentos. Similarly, in Op. 102 No. 2, Fournier (1947–48) performed at 133.74 bpm with 8 sliding portamentos, compared to Perényi (2001–02), who performed at 140.3 bpm with 7 sliding portamentos. These examples highlight that the decline in sliding portamento cannot be attributed to tempo changes alone.

\begin{table}[htbp]
\caption{Sliding Portamento Trends Across Sonatas}
\centering
\begin{tabular}{|c|c|c|c|}
\hline
\textbf{Sonata} & \textbf{Sliding (1930–2012)} & \textbf{\% Change} & \textbf{BPM Change} \\ \hline
Op. 5 No. 1   & 329 $\to$ 150 & -54.4\% & +15.5 \\ \hline
Op. 5 No. 2   & 382 $\to$ 149 & -61.1\% & +7.0  \\ \hline
Op. 69        & 524 $\to$ 342 & -34.7\% & +6.9  \\ \hline
Op. 102 No. 1 & 329 $\to$ 160 & -51.3\% & +8.6  \\ \hline
Op. 102 No. 2 & 141 $\to$ 80  & -43.3\% & +5.1  \\ \hline
\end{tabular}
\label{table:sliding_portamento_summary}
\end{table}

\begin{table}[htbp]
\caption{Silent Portamento Trends Across Sonatas}
\centering
\begin{tabular}{|c|c|c|c|}
\hline
\textbf{Sonata} & \textbf{Silent (1930–2012)} & \textbf{\% Change} & \textbf{Duration \%} \\ \hline
Op. 5 No. 1   & 164 $\to$ 116 & -29.1\% & -3.5\% \\ \hline
Op. 5 No. 2   & 51 $\to$ 45   & -11.1\% & +6.7\% \\ \hline
Op. 69        & 51 $\to$ 73   & +43.3\% & +0.15\% \\ \hline
Op. 102 No. 1 & 22 $\to$ 58   & +163.6\% & -4.7\% \\ \hline
Op. 102 No. 2 & 27 $\to$ 25   & -7.1\%  & -1.5\% \\ \hline
\end{tabular}
\label{table:silent_portamento_summary}
\end{table}

Silent portamento also shows no consistent correlation with tempo changes. In Op. 69, for example, Casals (1930–39) performed at 125.17 bpm with 1 silent portamento, whereas Isserlis (2012) performed at 127.01 bpm with 8 silent portamentos. The slight tempo increase of approximately 2 bpm does not account for the significant rise in silent portamento usage. These findings suggest that portamento usage is largely independent of tempo.

The data further reveals that sliding portamento usage tends to be higher in sonatas with slower rhythmic figures or lyrical qualities. For instance, Op. 69 recorded the highest total sliding portamento events (524), while Op. 102 No. 2, dominated by semiquaver passages, exhibited the lowest total sliding portamento count (141). However, even in works where tempo and rhythmic density would allow for more portamento, a consistent decline in sliding portamento usage is evident across the decades. Nevertheless, the available data does not establish a direct causal relationship between the HIP movement and the decline in portamento usage.

The decline in portamento usage, particularly sliding portamento, is evident across Beethoven’s cello sonatas and appears independent of tempo and performance duration. Silent portamento emerges as an alternative expressive tool in certain contexts, but its inconsistent application across the repertoire suggests that other factors, such as stylistic trends and cultural influences, play a more significant role.

\section{Portamento's Decline: A Discussion}

Musicologist Mark Katz identifies the ``phonograph effect'' as a significant factor contributing to the decline of portamento. He argues that early recording technologies, particularly the microphone, exaggerated the sound of slides to an extent deemed unappealing by both performers and listeners.~\cite{c9} As a result, musicians adapted their performance styles for recording purposes, intentionally minimising portamento to create cleaner, less intrusive recordings. This adaptation had lasting effects, influencing the overall approach to portamento and contributing to its decline after the 1970s. Writer David Blum similarly notes that Pau Casals advocated for a ``natural'' approach to glissando, viewing it as a tool for emotional expression when used judiciously, while violinist Leopold Auer criticised the overuse of portamento, considering it to be both inartistic and excessive.~\cite{c10}\cite{c11}  

\begin{figure}[h!]
  \caption{Blum’s example of Casals’ fingerings as extensions in J. S. Bach’s Suite for Violoncello Solo No. 4 BWV 1010, bar 1.}
  \includegraphics[width=1\textwidth]{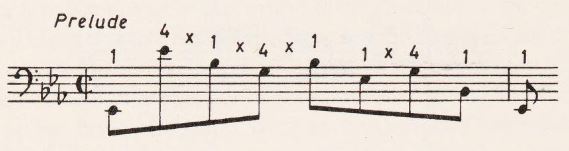}
\end{figure}

A notable development in performance practice during the 20th century was the emergence of new fingering techniques that reduced the need for portamento. Casals introduced over-stretched finger positions, which allowed players to reach notes without frequent shifting, thereby decreasing the necessity for slides. These techniques enabled faster tempos and cleaner transitions but, as David Blum observed, often restricted the expressive potential of the music.~\cite{c12} Violinist Otakar Ševčík's pedagogical exercises similarly promoted minimal shifts, emphasizing smooth transitions without portamento by encouraging efficient finger placements.~\cite{c13} These technical developments contributed to the reduction of portamento in both pedagogical and performance contexts.

\begin{figure}[htp!]

\begin{minipage}{.5\linewidth}
\centering
\subfloat[Casals' 1-2-3 fingering on the 6th position.]{\label{main:a}\includegraphics[scale=.5]{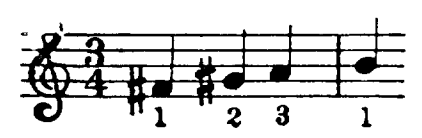}}
\end{minipage}%
\begin{minipage}{.5\linewidth}
\centering
\subfloat[Casals' 1-2-4 over-stretch finger extension on first position.]{\label{main:b}\includegraphics[scale=.5]{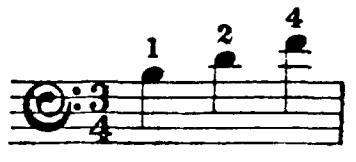}}
\end{minipage}\par\medskip
\centering
\subfloat[Casals' 1-2-3 fingerings applied in to a basic D major scale.]{\label{main:c}\includegraphics[scale=.5]{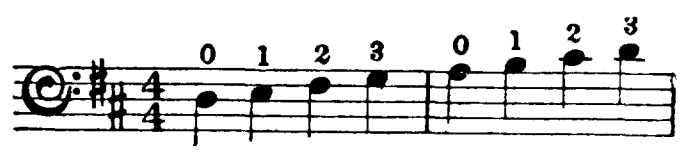}}

\caption[Cherniavsky's examples of what Casals' lizard technique would have looked like on paper.]{Cherniavsky's examples of what Casals' lizard technique would have looked like on paper.}
\label{fig:main}
\end{figure}

\bigskip

The impact of recording technology on performance style cannot be overstated. Microphones and recording processes amplified the subtle nuances of performance, including portamento, often rendering it overly prominent in recordings. As Katz explains, performers began to avoid or minimize portamento in recording sessions to achieve a more polished sound.~\cite{c14} Carl Flesch described how listening to recordings revealed unintentional portamento, prompting musicians to adjust their styles accordingly.~\cite{c15} This phenomenon was further echoed by violinist Zino Francescatti, who noted that performers tailored their approaches to ensure slides were subtle enough to suit recording technology.~\cite{c16} These adjustments highlight how recording technology not only documented performance but also reshaped its aesthetics.

\section{Conclusion}

Tempo in performances of Beethoven’s cello sonatas has shown a discernible increase when comparing data from 1930–1970 to 1970–2012. This trend is evident in recordings, with average tempos in sonatas like Op. 5 No. 1 rising from 135.78 BPM to 140.98 BPM, and Op. 102 No. 1 increasing from 135.52 BPM to 140.71 BPM. These changes reflect broader shifts in aesthetic preferences, favoring more energetic interpretations. Advances in instrument design, such as modern pianos with extended sustain and cellos with improved bows and strings, have facilitated faster tempos without compromising tonal quality. Pedagogical changes, including Casals’ fingering techniques, enabled performers to reduce hand shifts and play faster with greater precision. Additionally, the influence of recording technology likely reinforced this acceleration, as recordings emphasised clarity and minimised audible imperfections. However, while faster tempos have become more prevalent, lyrical works like Op. 69 remained relatively stable, suggesting that interpretative choices are also guided by expressive considerations.

The decline in sliding portamento, particularly after the mid-20th century, highlights the impact of technological and pedagogical shifts. Recording technologies exaggerated portamento’s audibility, leading performers to minimize its use for a cleaner sound. Pedagogical innovations, such as Casals’ lizard fingerings, further reduced the need for expressive shifts, favoring technical precision. Although silent portamento emerged as an alternative, its inconsistent use highlights its limited capacity to replicate the expressive depth of sliding portamento.

A major challenge of this study lies in analysing portamento and tempo within polyphonic textures, where overlapping voices obscure key details in recordings. Beethoven’s contrapuntal passages often demand nuanced interpretative decisions that are difficult to isolate in such dense structures. Future research could address this limitation by employing audio separation tools like Spleeter to isolate individual instrumental lines, enabling more precise analysis of portamento and tempo across polyphonic contexts. This, combined with machine learning techniques and expanded methodologies such as performer interviews and live analyses, could further enhance understanding of how historical intent interacts with modern practices.

I hope this study encourages further exploration into the intersection of tempo, portamento, and performance practice in Beethoven’s works. Please feel free to reach out with any questions.



\begin{thebibliography}{99}
  
    \bibitem{c1} Dr. Kalischer, \textit{Beethoven's Letters,} trans. J. S. Shedlock, Dover Publications, 1972, letter 416.

    \bibitem{c2} Lawrence Talbot, \textit{A Note on Beethoven's Metronome}, California: University of California Press, 1971, p. 328.

    \bibitem{c3} Sture Forsen et al., \textit{Was Something Wrong with Beethoven's Metronome?}, Notices of the AMS, 2013, pp. 1150-51. 

    \bibitem{c4} Iñaki Ucar and Almudena Martín-Castro, \textit{Conductor's Tempo Choices Shed Light Over Beethoven's Metronome}, PLoS ONE 15(12), 2020, p. 6.

    \bibitem{c5} Steven Isserlis, \textit{Letters with Ignasi}, March 28, 2022.

    \bibitem{c6} Tilman Skowroneck, \textit{Beethoven's Erard Piano: Its Influence on His Compositions and on Viennese Fortepiano Building}, Early Music, Vol. 30, No. 4, Oxford University Press, 2002, pp. 523-34.

     \bibitem{c7} Marten Noorduin, \textit{Beethoven's Tempo Indications}, A thesis submitted to the University of Manchester for the degree of Doctor of Philosophy in the Faculty of Humanities, 2016, p. 14. \url{https://www.escholar.manchester.ac.uk/uk-ac-man-scw:302884}

     \bibitem{c8} Rudolf, Kolisch, \textit{Tempo and Character in Beethoven's Music}, The Musical Quarterly, Spring, 1993, Vol. 77, No. 1, pp. 90-131.

     \bibitem{c9} Mark Katz, \textit{Portamento And The Phonograph Effect}, Journal of Musicological Research, Vol. 25, 2006, pp. 211–32.

     \bibitem{c10} David Blum, \textit{Casals and the Art of Interpretation}, University of California Press, 1977, pp. 125–31.

     \bibitem{c11} Leopold Auer, \textit{My Long Life in Music}, Frederick A. Stokes, 1923, p. 34.

     \bibitem{c12} David Blum, \textit{Casals and the Art of Interpretation}, University of California Press, 1977, pp. 130–31.

     \bibitem{c13} Otakar Ševčík, \textit{School of Violin Technique, Op.1}, G. Schirmer, 1905.

     \bibitem{c14} Mark Katz, \textit{Portamento And The Phonograph Effect}, Journal of Musicological Research, Vol. 25, 2006, pp. 225–27.

     \bibitem{c15} Carl Flesch, \textit{Art of Violin Playing}, Carl Fischer, 2008, p. 85.

     \bibitem{c16} Zino Francescatti, \textit{With the Artists}, John Markert, 1955, p. 18.

    
\end{thebibliography}
\end{document}